\documentclass[pra,twocolumn,showkeywords,longbibliography,superscriptaddress,10pt]{revtex4-1}
\usepackage{graphicx}
\usepackage{dcolumn}
\usepackage{color}
\usepackage{times}
\usepackage{bm}
\usepackage{amssymb}
\usepackage{amsmath}
\usepackage{epsfig}
\usepackage{epstopdf}
\usepackage{dsfont}
\usepackage{subfigure}
\usepackage{tikz}
\usepackage[colorlinks, citecolor=blue, linkcolor=blue,urlcolor=blue]{hyperref}
\usepackage[mathscr]{euscript}
\makeatletter
\renewcommand{\maketag@@@}[1]{\hbox{\m@th\normalsize\normalfont#1}}
\makeatother
\begin{document}
	\renewcommand{\thefootnote}{\fnsymbol {footnote}}
	
	\title{{Environment-mediated} entropic uncertainty in charging quantum batteries}
	%Effect of Hawking radiation on entropic uncertainty relations in the presence of quantum memory
	
	\author{Meng-Long Song}
	\affiliation{School of Physics \& Optoelectronic Engineering, Anhui University, Hefei
		230601,  People's Republic of China}
	
	\author{Li-Juan Li}
	\affiliation{School of Physics \& Optoelectronic  Engineering, Anhui University, Hefei 230601, People's Republic of China}
	
	\author{Xue-Ke Song} \email{songxk@ahu.edu.cn}
	\affiliation{School of Physics \& Optoelectronic Engineering, Anhui University, Hefei 230601,  People's Republic of China}
	
	\author{Liu Ye}
	\affiliation{School of Physics \& Optoelectronic Engineering, Anhui University, Hefei 230601,  People's Republic of China}
	
	\author{Dong Wang} \email{dwang@ahu.edu.cn}
	\affiliation{School of Physics \& Optoelectronic Engineering, Anhui University, Hefei
		230601,  People's Republic of China}

	%\author{Liu Ye} %\email{yeliu@ahu.edu.cn}
	%\affiliation{School of Physics \& Optoelectronic Engineering, Anhui University, Hefei
		%230601,  People's Republic of China}

	\date{\today}
	
	\begin{abstract}
		{{We studied the dynamics of entropic uncertainty in Markovian and non-Markovian systems during the charging of open quantum batteries (QBs) mediated by a common dissipation environment. In the non-Markovian regime, the battery is almost fully charged efficiently, and the strong non-Markovian property is beneficial for improving the charging power.
				In addition, the results show that the energy storage is closely related to the couplings of the charger-reservoir and battery-reservoir; that is, the stronger coupling of a charger-reservoir improves energy storage.
				In particular, entanglement is required to obtain the most stored energy and is accompanied by the least tight entropic bound. Interestingly, it was found that the tightness of the entropic bound can be considered a good indicator of the energy transfer in different charging processes, and the complete energy transfer always corresponds to the tightest entropic bound. Our results provide insight into the optimal charging efficiency of QBs during  practical charging.}}
		
	\end{abstract}
	
	\maketitle

	\section{Introduction}
	The concept of quantum battery (QB) was {proposed} by Alicki and Fannes \cite{l7ah1}, {who investigated} the extractability of the amount of work done by a quantum system {during} unitary evolution. It is a device used to temporarily store energy. In recent years, a series of studies {have been conducted} on energy changes in quantum batteries (QBs) \cite{l7ah1,l9,l12,l8ab12,l10ab3ah29,l11ab2}. %\textcolor[rgb]{1.00,0.00,0.00}{}
	{In practice,} both quantum and classical batteries focus on the efficiency with which energy (in the case of the battery) is transferred during charging and discharging {to extract} as much energy as possible from the battery.  QB considered an open system because of its interaction with the surrounding environment. This leads to the leakage of energy {from the} QB into the environment. {Thus} far, many scholars have proposed various schemes for studying energy transfer efficiency by quantum {effects} \cite{l14,l15,l13ab13}.
	{In principle, the characterizations for improving the performance of the QB during charging are reflected by higher energy storage and the smaller optimal charging time (i.e., the shortest time $\lambda t_{s}$ to reach the peak energy storage)
		\cite{l11ab2,l10ab3ah29,w1,w2,b4,b6,b7,b8}. Therefore, it is necessary to pursue a better charging process by tuning the control parameters in the transfer process.}

	{In addition}, the uncertainty relation was originally proposed by Heisenberg, {where} we cannot accurately predict the measurement outcomes with respect to two arbitrary incompatible observables \cite{1}. Heisenberg's uncertainty principle is regarded as {a cornerstone} of quantum mechanics. Subsequently, Kennard \cite{2} and Robertson \cite{3} proposed {an expression of} the uncertainty principle in the form of standard deviation: $\Delta {\hat{P}_1} \cdot \Delta {\hat{P}_2} \ge \frac{1}{2}| \langle {[{\hat{P}_1},{\hat{P}_2}]}  \rangle |$, where $\Delta (\bullet)$ represents the standard deviation, and $ \langle \circ\rangle$ in the lower bound represents the corresponding expected value. Note that this relation {leads} to a trivial result, that is, the lower bound will become zero when the system's state is prepared in one of the two observables' eigenstates.  On this premise, Deutsch introduced information entropy to describe the uncertainty principle and {present the} entropic uncertainty relation (EUR) \cite{4}: $H( {\hat{{P_1}}} ) + H( {\hat{{P_2}}} ) \ge {\log _2}{( {\frac{2}{{1 + \sqrt c }}} )^2},$ where $c = {\max _{ij}}|\left\langle {{{P_{1i}}}}.
	\mathrel{\left | {\vphantom {{{P_{1i}}} {{P_{2j}}}}}
		\right. \kern-\nulldelimiterspace}
	{{{P_{2j}}}} \right\rangle {|^2}$, $\left| {{P_{1i}}} \right\rangle $ and $\left| {{P_{2j}}} \right\rangle $ being with the eigenstates of the measurement operator ${\hat{{P_1}}}$ ${\hat{{P_2}}}$, and the Shannon entropy $H ( {{\hat{P}_k}}  ) =  - \sum\nolimits_i {{p_{ki}}} {\log _2}{p_{ki}} \left( {k = 1,2} \right)$  with ${p_{ki}} = \left\langle {{P_{ki}}} \right|{ \hat \rho } \left| {{P_{ki}}} \right\rangle $ being the probability of obtaining the $i$th  measurement result. Later, Kraus \cite{5} and Maassen and Uffink \cite{6}
	improved Deutsch's {results}. Notably, {this relationship} is suitable for single-particle systems. Thus, one may ask how to express the uncertainty relation.
	If the measured particle is correlated with another. Berta {\it et al.} proposed a new form of uncertainty relation called quantum memory-assisted entropic uncertainty relation (QMA-EUR) \cite{7,m,x}. {Their investigation focuses on a two-measurement case.} {Later on, Liu $et$ $al.$  expanded Berta {\it et al.}'s version to the case of multiple measurements, and put forward the general expression of a QMA-EUR related to multiple measurements as \cite{liu}
		\begin{align}
			\sum\limits_{x = 1}^N {S\left( {{{\hat P}_x}\left| B \right.} \right)}  \ge  - {\log _2}\left( b \right) + \left( {N - 1} \right)S\left( {A\left| B \right.} \right),
			\label{Eq.1}
		\end{align}
		where $S ( {A|B}  ) = S ( {{{\hat \rho }_{AB}}}  ) - S\left( {{{\hat \rho }_B}} \right)$ is the von Neumann conditional entropy \cite{8}, and the  von Neumann entropy $S\left( {\hat \rho } \right) =  - tr\left( {\hat \rho {{\log }_2}\hat \rho } \right)$. Specifically, $b = \mathop {\max }\limits_{{i_N}} \left\{ {\sum\limits_{{i_2} \sim {i_{N - 1}}} {\mathop {\max }\limits_{{i_1}} } \left[ {c\left( {u_{{i_1}}^1,u_{{i_2}}^2} \right)} \right]\prod\limits_{x = 2}^{N - 1} {\left[ {c\left( {u_{{i_x}}^x,u_{{i_{x + 1}}}^{x + 1}} \right)} \right]} } \right\}$ and $u_{i_x}^x$ is the $i$th eigenvector of the operator ${{{\hat P}_x}}$.}
	
	{Recently}, some scholars have studied the change {in} entanglement in the QB, \cite{x18} and the relationship between energy density and entanglement \cite{x18+} during quantum battery charging, which indicates that entanglement plays an important role in energy transfer during {the} charging process. {In addition}, the connection between entanglement and uncertainty has been disclosed \cite{ee}.
	With this in mind, the tightness of the lower bound on {the} entropic uncertainty relation plays {a role} during the charging of quantum batteries. {Motivated by this, we first investigate the optimization of the charging process by different coupling regimes: Markovian and non-Markovian regimes. When the battery is stable, we look for more stored energy by investigating the qubit-environment coupling and the initial entanglement of the charger and battery. We found that a stronger charger-environment coupling is beneficial for obtaining more stored energy. Moreover, we study the variation in  the entropic uncertainty relation (the tightness of the entropic bound) and the relation with energy storage in charging QBs. Remarkably, it was found that the tightness of the entropic bound can be considered an indicator of the energy transfer, and the complete energy transfer is always accompanied by the tightest entropic bound. That is, the tightness of the uncertainty bound can reflect the charging efficiency in the QB.}

	The structure of this paper is as follows. {In} Section II, charging model of {an} open quantum battery system is reviewed. In Section III, {the time evolution of stored energy and power in different coupling regimes is studied in the model.} In Section IV, {we investigate the steady-state energy of the QB and its relation to the tightness of the lower bounds on entropic uncertainty.} {Furthermore, we investigate the tightness at different energy transfer rates (the proportion of energy transferred from the charger to the battery) in the charging process.} Finally, we {conclude our study} with a concise summary in Section V.

	\section{charging model}
	We consider a quantum battery charging model consisting of a quantum charger $A$ and a quantum battery $B$, which {is} coupled to a common zero-temperature bosonic reservoir, without coupling between $A$ and $B$. Each cell of $A$ and $B$ is a two-level system with the same transition frequency, ${\omega _A} = {\omega _B} = {\omega _0}$, and the ground and excitation states are $\left| {{g}} \right\rangle $ and $\left| e \right\rangle $, respectively. The Hamiltonian $H$ of the {entire} system is composed {of} ${H_0}$ and ${H_{{\mathop{\rm int}} }}$ \cite{b23,b24,b}
	\begin{figure}
		\begin{minipage}{0.45\textwidth}
			\centering
			\subfigure{\includegraphics[width=8cm]{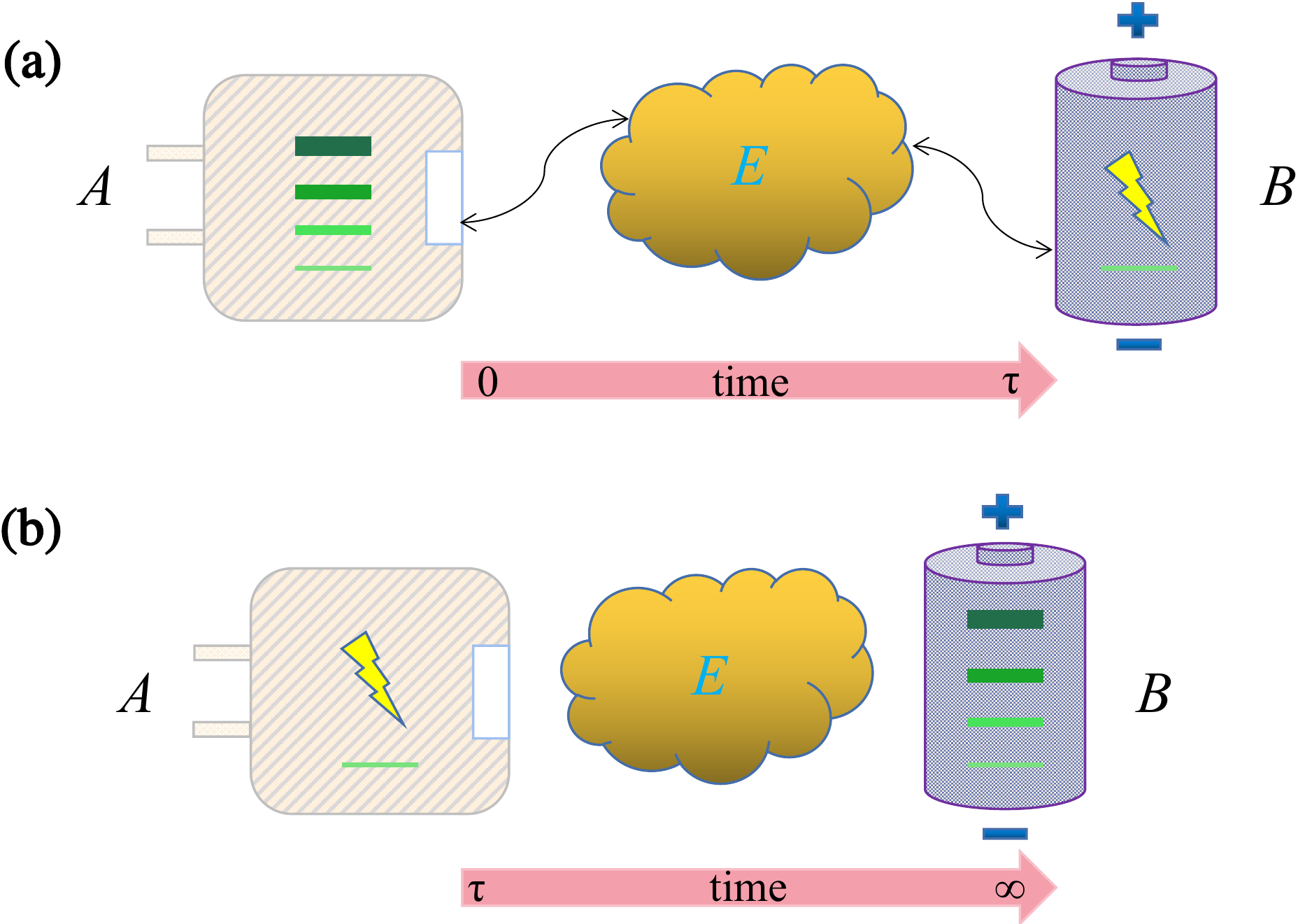}}
		\end{minipage}\hfill
		\caption{ (a)  Within time $t \in \left[ {0,\tau } \right]$, the charger ($A$) and battery ($B$) interacts with the environment ($E$). (b) When $t > \tau$, at the end of the charging process, the battery is in the steady-state and its energy is conserved.}
		\label{f1}
	\end{figure}

	\begin{align}
		H = {H_0} + {H_{{\mathop{\rm int}} }},
		\label{Eq.2}
	\end{align}
	where
	\begin{align}
		{H_0} = \sum\limits_{i = A,B} {{\omega _0}} \sigma _i^ + \sigma _i^ -  + \sum\limits_k {{\omega _k}c_k^\dag {c_k}} ,
		\label{Eq.3}
	\end{align}
	and
	\begin{align}
		{H_{{\mathop{\rm int}} }} = \left( {{\beta _1}\sigma _A^ +  + {\beta _2}\sigma _B^ + } \right)\sum\limits_k {{g_k}{c_k}}  + \left( {{\beta _1}\sigma _A^ -  + {\beta _2}\sigma _B^ - } \right)\sum\limits_k {{g_k}c_k^\dag } .
		\label{Eq.4}
	\end{align}
	${H_0}$ represents the free Hamiltonian of the qubit and the reservoir, where $\sigma _i^ + $ and $\sigma _i^ - $ are the Pauli rising and lowering operators of the qubit, ${\omega _k}$ and $c_k^\dag $(${c_k}$) are the frequency and creation(annihilation) operators of the $k$th mode of the field, respectively. ${H_{{\mathop{\rm int}} }}$ denotes the interaction between the two-qubit system and reservoir. ${\beta _1}$ and ${\beta _2}$ are defined as the interaction strengths between qubits $A$ and $B$ and {the} reservoir, respectively, {and} are dimensionless real parameters.
	And ${g_k}{\beta _1}$ and ${g_k}{\beta _2}$ are the coupling constants between {the} qubits and reservoir. {Then, the} collective coupling constant ${\beta _T}=\sqrt {\beta _1^2 + \beta _2^2} $ and {the} relative interaction strength ${\zeta _i}={{{\beta _i}} \mathord{\left/{\vphantom {{{\beta _i}} {{\beta _T}}}} \right.\kern-\nulldelimiterspace} {{\beta _T}}} \left( {i = 1,2} \right)$. {It should be noted} that different effective {couplings} of $A$ and $B$ with {a} reservoir can be obtained under different conditions {for} ${{\beta _1}}$ and ${{\beta _2}}$.
	\begin{figure*}[htbp]
		\centering
		\subfigure{\includegraphics[height=6cm]{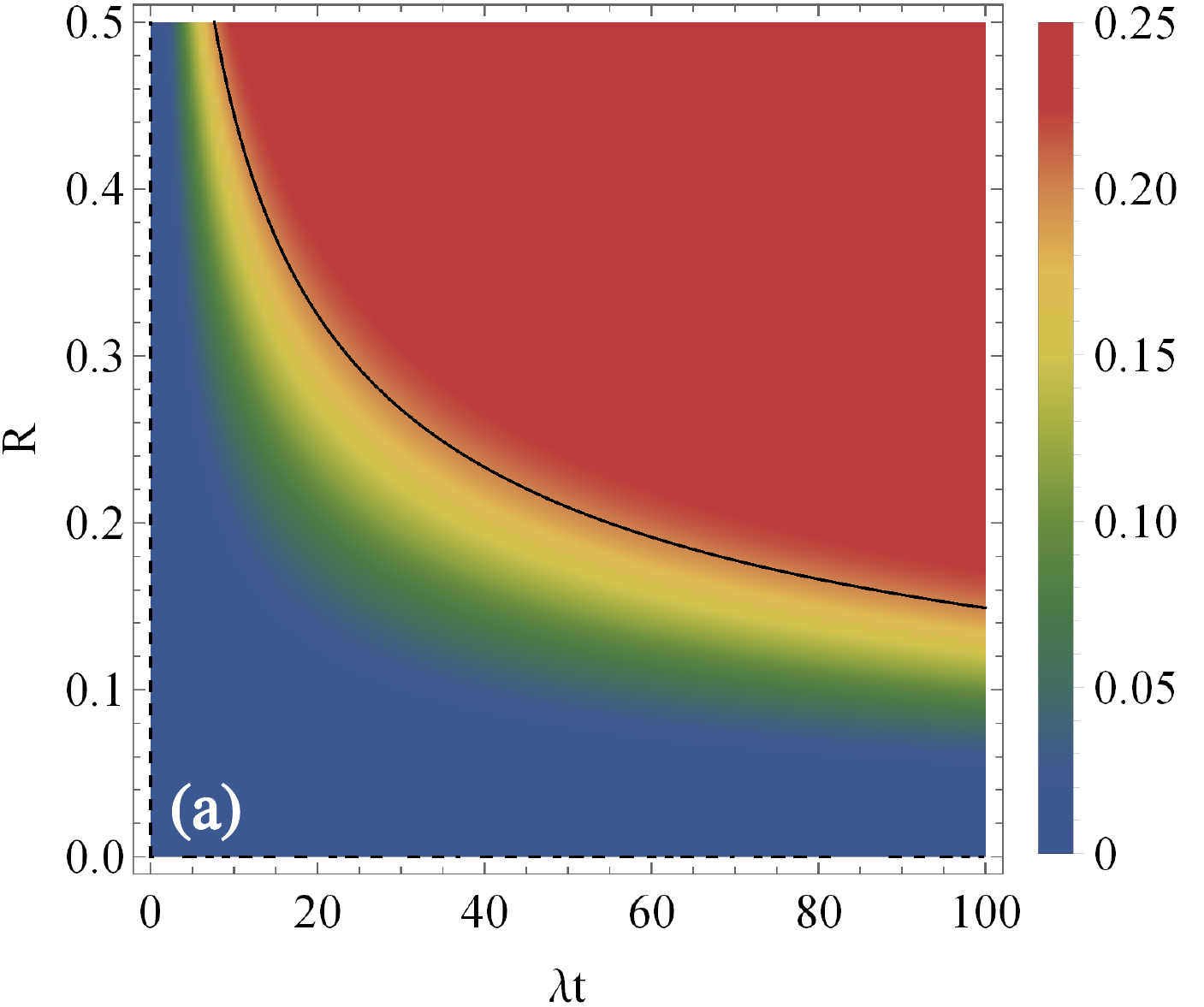}}
		\hspace{0.5cm}
		\subfigure{\includegraphics[height=6cm]{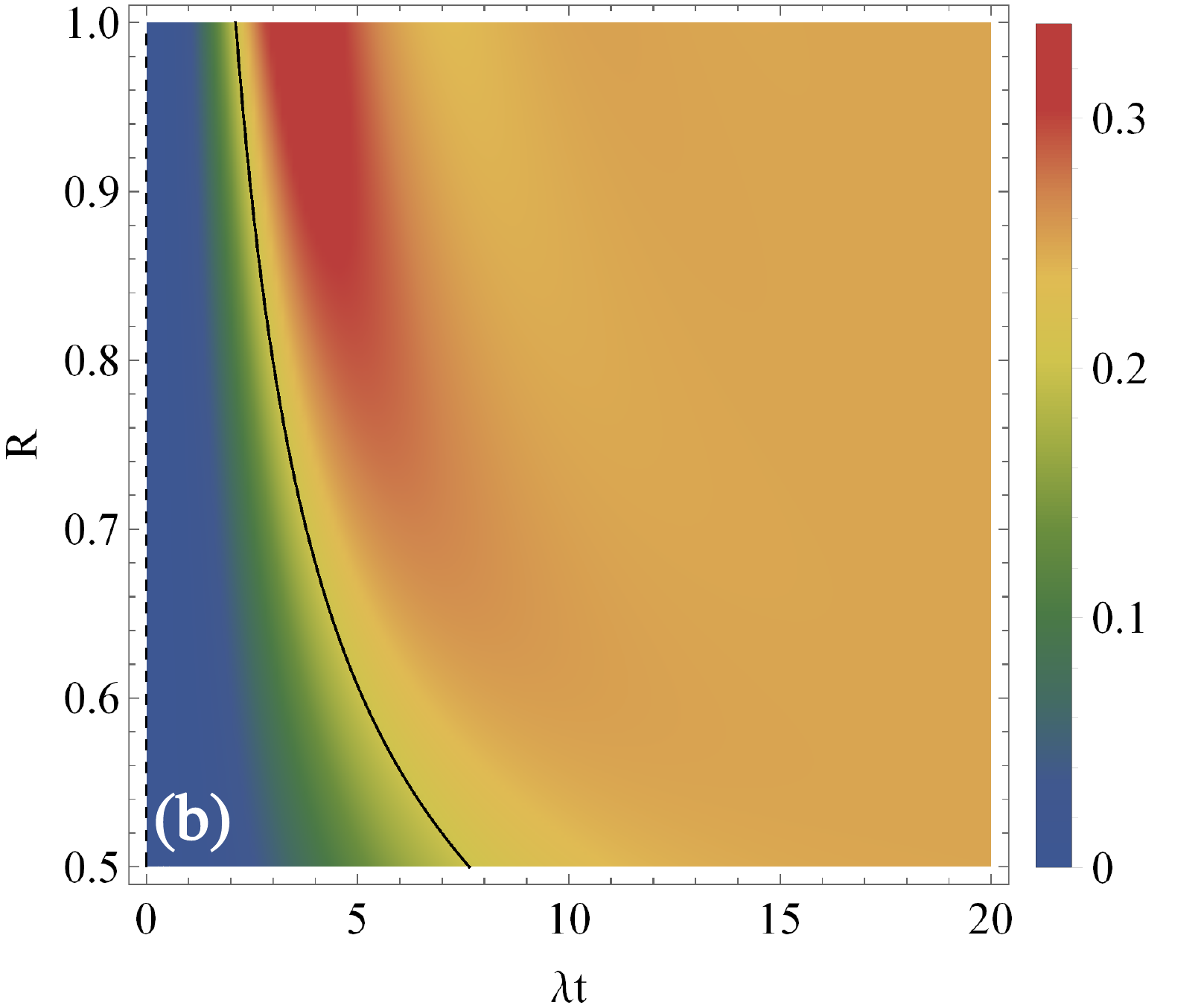}}
		\\
		\subfigure{\includegraphics[height=6cm]{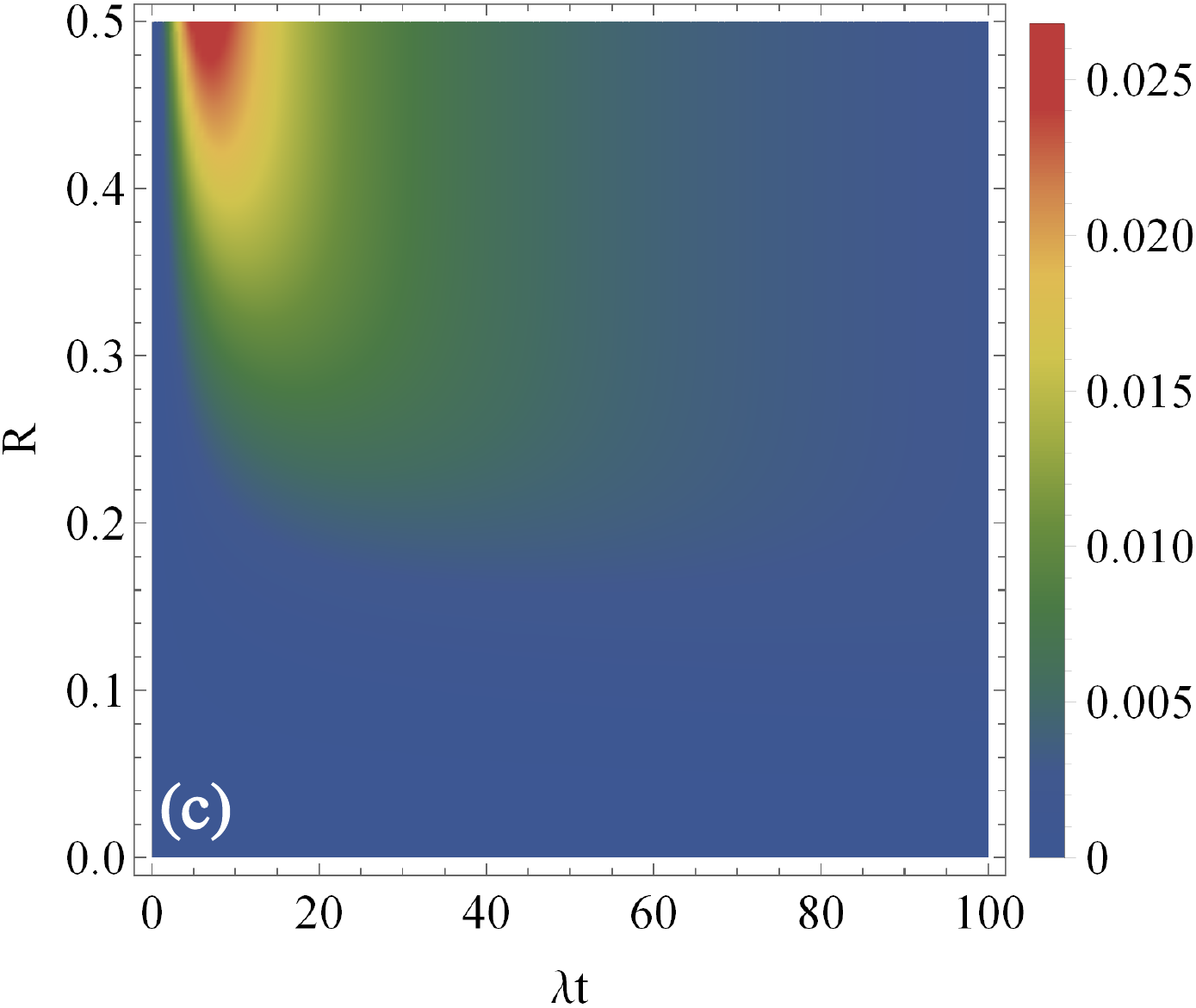}}
		\hspace{0.5cm}
		\subfigure{\includegraphics[height=6cm]{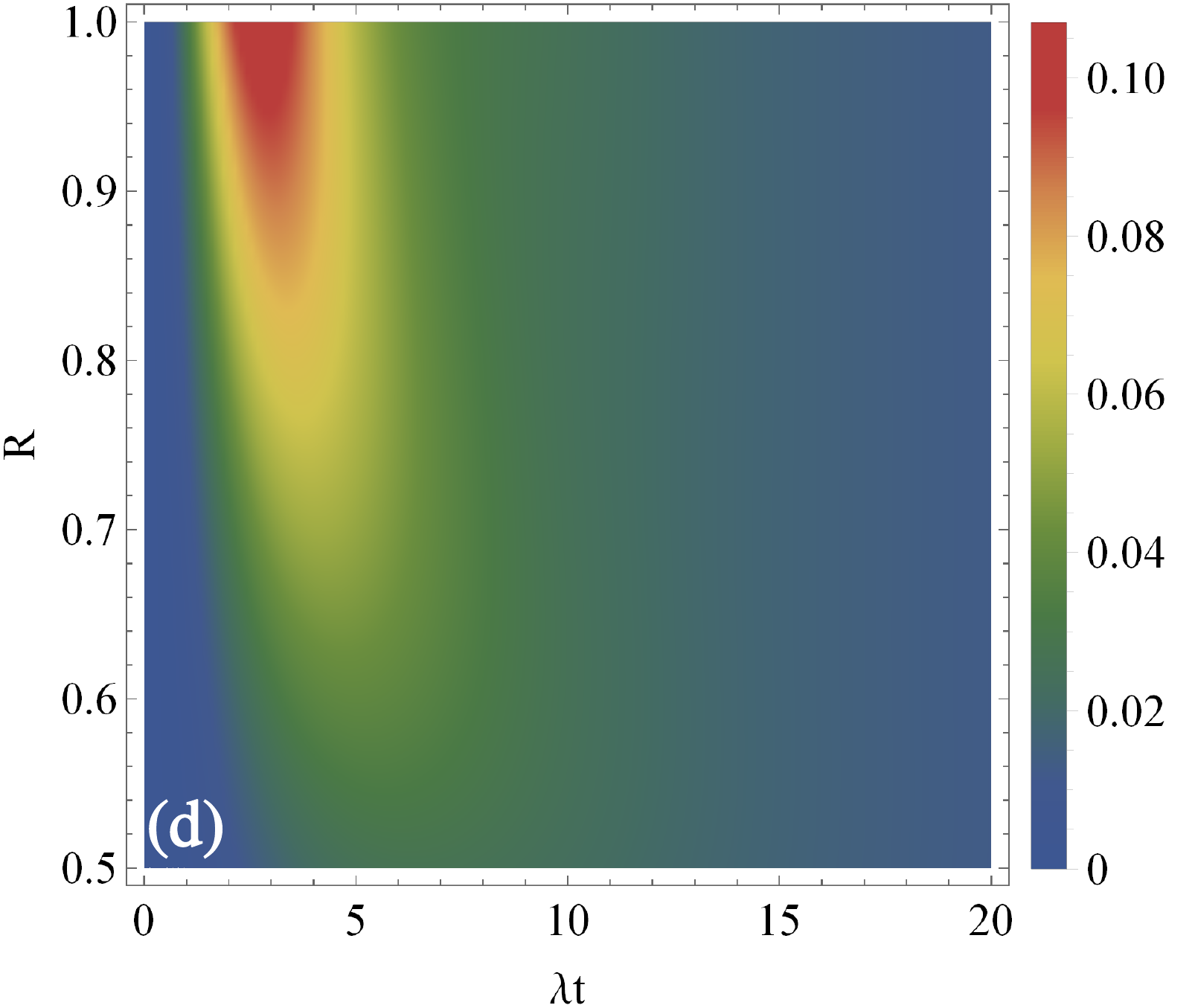}}
		\caption{{Markovian charging process. The stored energy of battery $\Delta {E_B}\left( t \right)/{\omega _0}$ and instantaneous power $P/{\omega _0}$ as function of the dimensionless quantity $\lambda t$ and coupling strength $R$  are
				plotted in (a), (b) and (c), (d), respectively. We choose the initial separable states with $\theta  = 0.5\pi$, and the same interactions of charger-reservoir and battery-reservoir (${\zeta _1} = {\zeta _2} =  \frac{1}{{\sqrt 2 }}$). Graphs (a) and (c): $0\leq R\leq 0.5$; Graphs (b) and (d): $0.5 \leq R\leq 1$.}}
		\label{f2}
	\end{figure*}
	In addition, ${H_{{\mathop{\rm int}} }}$ exists within the time interval: $t \in \left[ {0,\tau } \right]$ of the charging process. Outside this interval, there is no interaction between $A$($B$) and the environment, {furthermore,} there is no interaction between $A$ and $B$. Therefore, this is {the} wireless charging process of QB in {an environmental} medium \cite{b}. Note that, this charging model corresponds to physically feasible   experimental settings by the all-optics platform \cite{x19,x20,b}.
	
	Under the limitation that the two-qubit system and reservoir are separable states at the beginning, we consider the case in which there is only one excitation state in the two-qubit system and the reservoir is in a vacuum state.  The initial state of the {entire} system is
	\begin{align}
		\left| {\varphi \left( 0 \right)} \right\rangle  = {\eta _{01}}{\left| e \right\rangle _A}{\left| {{g}} \right\rangle _B}{\left| 0 \right\rangle _{\rm X}} + {\eta _{02}}{\left| {{g}} \right\rangle _A}{\left| e \right\rangle _B}{\left| 0 \right\rangle _{\rm X}},
		\label{Eq.5}
	\end{align}
	where, ${\eta _{01}}$ and ${\eta _{02}}$ are {the probabilities} {amplitudes} and ${\left| 0 \right\rangle _{\rm X}}$ is the vacuum state of the reservoir. The time evolution of a single excited state can be obtained as
	\begin{align}
		\left| {\varphi \left( t \right)} \right\rangle  &= {\eta _1}\left( t \right){\left| e \right\rangle _A}{\left| {{g}} \right\rangle _B}{\left| 0 \right\rangle _{\rm X}} + {\eta _2}\left( t \right){\left| {{g}} \right\rangle _A}{\left| e \right\rangle _B}{\left| 0 \right\rangle _{\rm X}}\nonumber\\
		&+ \sum\limits_k {{\eta _k}\left( t \right)} {\left| {{g}} \right\rangle _A}{\left| {{g}} \right\rangle _B}{\left| 1_k \right\rangle _{\rm X}},
		\label{Eq.6}
	\end{align}
	where, ${\eta _1}\left( t \right)$ and ${\eta _2}\left( t \right)$ are probability amplitudes. We consider the case {in which} the environment is regarded as an electromagnetic field in a lossless cavity, and the Lorentzian form of the spectral density of the cavity field is \cite{b26}
	\begin{align}
		J\left( \omega  \right) = \frac{{{\nu ^2}\lambda }}{{\pi {{\left( {\omega  - {\omega _0}} \right)}^2} - \pi {\lambda ^2}}},
		\label{Eq.7}
	\end{align}
	where, $\lambda $ is the spectrum width {and} ${1 \mathord{\left/
			{\vphantom {1 \lambda }} \right.
			\kern-\nulldelimiterspace} \lambda }$ is the correlation time of the reservoir, and $\nu $ is the effective coupling strength satisfying $L = \nu {\beta _T}$ {where} $L$ is the vacuum Rabi frequency. Then we define a dimensionless real number ${R} = L/{\lambda }$, where strong coupling has $ {{R} \gg 1} $ and weak coupling with ${{R} \ll 1} $ \cite{b191,b192,b193,b194,b195,b196,b197}. In this case, Markovian evolution occurs under weak coupling, and non-Markovian dynamics and memory effects {occur} under strong coupling. And then the analytic solution for ${\eta _i}\left( t \right) \left( {i = 1,2} \right)$ comes {in} the following form \cite{b23,b24,b}
	\begin{align}
		{\eta _1}\left( t \right) = \left[ {\zeta _2^2 + \zeta _1^2\mu \left( t \right)} \right]{\eta _{01}} - {\zeta _1}{\zeta _2}\left[ {1 - \mu \left( t \right)} \right]{\eta _{02}},
		\label{Eq.8}
	\end{align}
	and
	\begin{align}
		{\eta _2}\left( t \right) = \left[ {\zeta _1^2 + \zeta _2^2\mu \left( t \right)} \right]{\eta _{02}} - {\zeta _1}{\zeta _2}\left[ {1 - \mu \left( t \right)} \right]{\eta _{01}},
		\label{Eq.9}
	\end{align}
	where the Lorentzian form is $\mu \left( t \right) = {e^{ - \frac{{\lambda t}}{2}}}\left[ {\cosh \left( {\frac{{\kappa t}}{2}} \right) + \frac{\lambda }{\kappa }\sinh \left( {\frac{{\kappa t}}{2}} \right)} \right]$ {and} $\kappa  = \sqrt {{\lambda ^2} - 4{L^2}} $ \cite{b23,b24,b}. {Finally, we have
		\begin{align}
			\mu \left( t \right) = \left\{ {\begin{array}{*{20}{c}}
					{{e^{ - \frac{{\lambda t}}{2}}}\left[ {\cosh \left( x \right) + \sinh \left( x \right)/{\chi _1}} \right],}&{0.5 \ge R \ge 0}\\
					{{e^{ - \frac{{\lambda t}}{2}}}\left[ {\cos \left( y \right) + \sin \left( y \right)/{\chi _2}} \right],}&{R \ge 0.5}
			\end{array}} \right.
			\label{Eq.10}
		\end{align}
		where $x = {\chi _1}\sqrt {{\lambda ^2}{t^2}} /2$, $y = {\chi _2}\sqrt {{\lambda ^2}{t^2}} /2$ and ${\chi _1} = \sqrt {1 - 4{R^2}} $, ${\chi _2} = \sqrt {4{R^2} - 1} $.}

	\section{Performance of quantum batteries with different coupling regimes}
	
	During the charging process of {the} quantum batteries, {the coupling strength between the qubits (charger and battery) and the reservoir is considered to be an important element in determining excellent charging performance.}
	%in order to observe the evolution process of entropic uncertainty relation of the system under different initial states and coupling strength, we resort to a pair of incompatible measurements ($\sigma _x,\sigma _z$).
	From Eqs. (\ref{Eq.5}) and (\ref{Eq.6}),  the reduced density matrices can be {expressed as:}
	\begin{align}
		\rho _{AB}{\left( 0 \right)} &= |{\eta _{01}}{|^2}{\left| {e{{g}}} \right\rangle _{AB}}\left\langle {e{{g}}} \right| + |{\eta _{02}}{|^2}{\left| {{{g}}e} \right\rangle _{AB}}\left\langle {{{g}}e} \right|\nonumber\\
		&+ {\eta _{01}}\eta _{02}^*{\left| {e{{g}}} \right\rangle _{AB}}\left\langle {{{g}}e} \right| + \eta _{01}^*{\eta _{02}}{\left| {{{g}}e} \right\rangle _{AB}}\left\langle {e{{g}}} \right|,
		\label{Eq.11}
	\end{align}
	and
	\begin{align}
		{\rho _{AB}}\left( t \right) &= |{\eta _1}\left( t \right){|^2}{\left| {eg} \right\rangle _{AB}}\langle eg| + |{\eta _2}\left( t \right){|^2}{\left| {ge} \right\rangle _{AB}}\langle ge|\nonumber\\
		&+ {\eta _1}\left( t \right){\eta _2}{\left( t \right)^*}{\left| {eg} \right\rangle _{AB}}\langle ge| + {\eta _1}{\left( t \right)^*}{\eta _2}\left( t \right){\left| {ge} \right\rangle _{AB}}\langle eg|\nonumber\\
		&+ \left[ {1 - |{\eta _1}\left( t \right){|^2} - |{\eta _2}\left( t \right){|^2}} \right]{\left| {gg} \right\rangle _{AB}}\langle gg|
		\label{Eq.12}
	\end{align}
	of the two-qubit system. {Note that $\eta_{i}(t)=\eta_{i}(t)^{*}, (i=1,2)$.} The {reduced-density matrices} $\rho_A {\left( t \right)}$ and $\rho _B{\left( t \right)}$ of $A$ and $B$ can also be obtained. {Then we take ${\eta _{01}} = \sin \theta  \cdot{e^{i\phi }}$ and ${\eta _{02}} = \cos \theta $.}
	According to Eq. (\ref{Eq.12}) and {the} reduced density matrix $\rho _B{\left( t \right)}$, the energy of the battery at time $t$ is represented by ${{ E}_B}\left( t \right) = tr\left[ {{H_B} \rho_B {{\left( t \right)}}} \right]$. Hence, the change in battery energy during charging can be quantified by
	\begin{align}
		\Delta {E_B}\left( t \right) &= {E_B}\left( t \right) - {E_B}\left( 0 \right)\nonumber \\
		&= {\omega _0}\left[ {|{\eta _2}\left( t \right){|^2} - \cos^{2} {\theta }} \right].
		\label{Eq.13}
	\end{align}
	{Another effective parameter in the charging process is the charging power of the QB. In this case, the instantaneous power of the charging process is:
		\begin{align}
			{P} = \frac{{ {E_B}(t)}}{t}.
			\label{Eq.14}
		\end{align}}

	\subsection{{Markovian dynamics}}
	
	\begin{figure*}
		\begin{minipage}{1\textwidth}
			\centering
			\subfigure{\includegraphics[width=7cm]{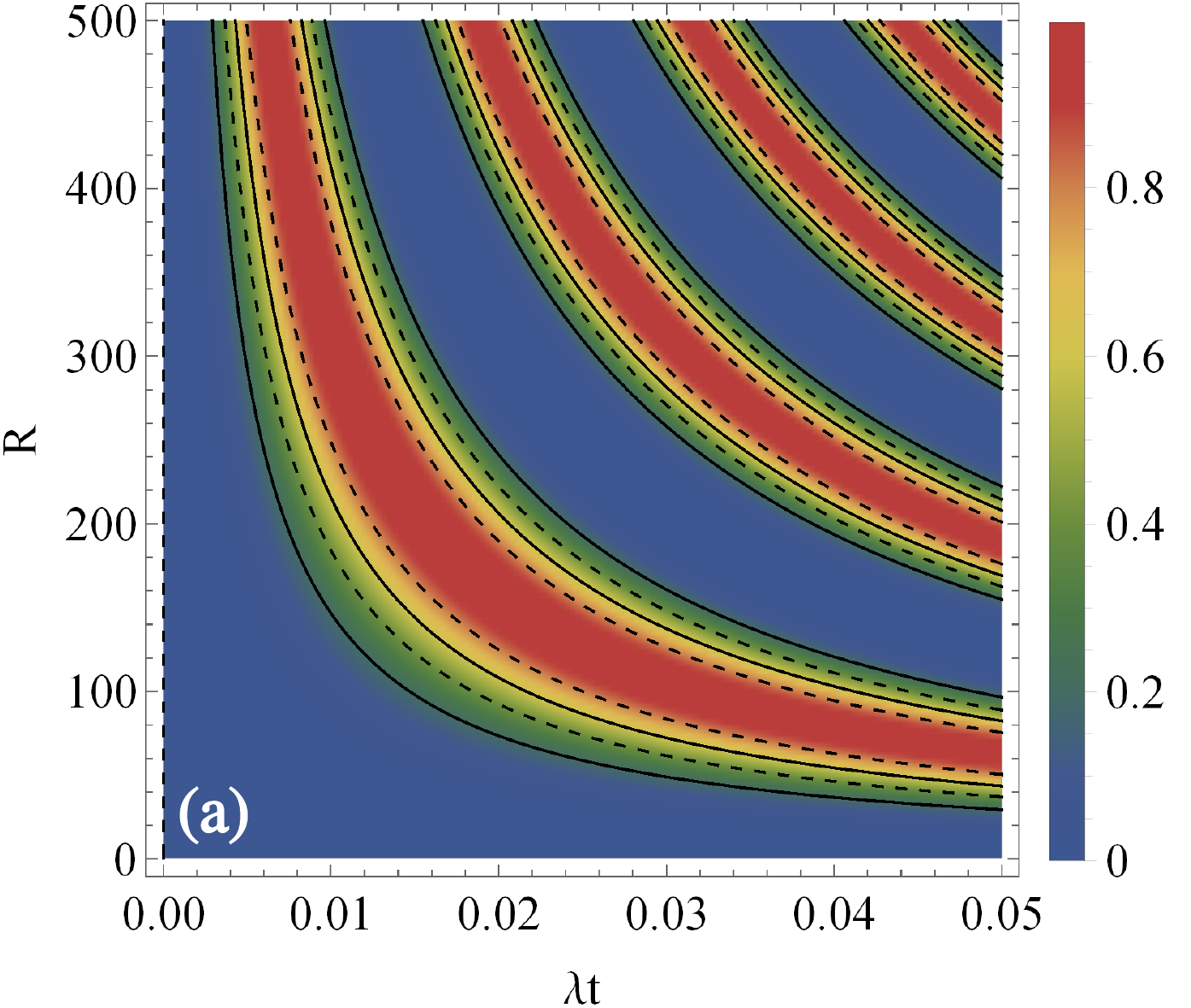}} \ \ \ \ \
			\subfigure{\includegraphics[width=6.8cm]{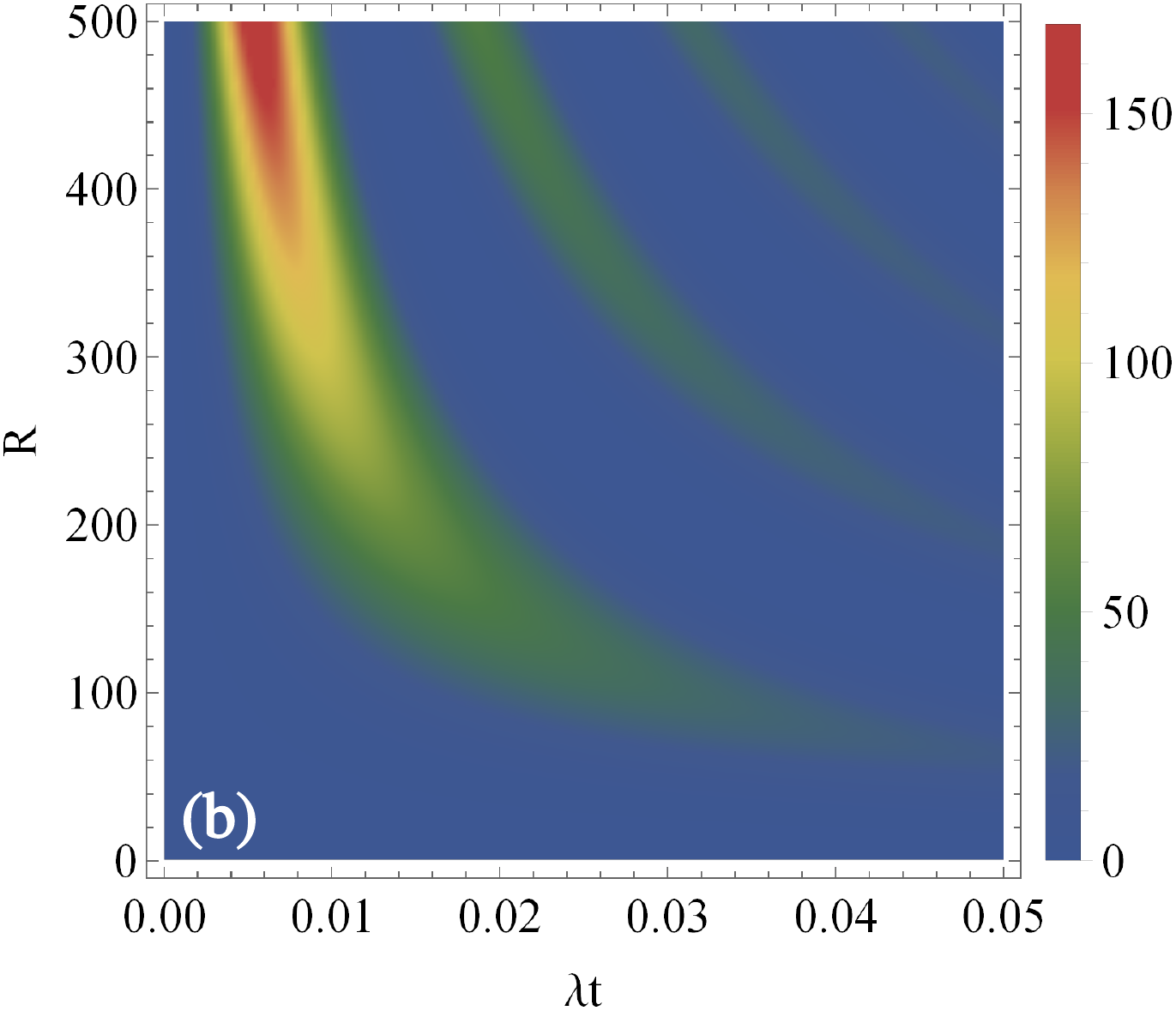}}
		\end{minipage}\hfill
		\caption{{In the Non-Markovian charging process the stored energy of the battery $\Delta {E_B}\left( t \right)/{\omega _0}$ and instantaneous power $P/{\omega _0}$ as function of the dimensionless quantity $\lambda t$ and coupling strength $R$  are plotted in Graphs (a) and (b), respectively. $1\leq R\leq 500$ is set and all other parameters are the same as those in Fig. \ref{f2}.}}
		\label{f3}
	\end{figure*}
	{In the weakly coupled regime, the general behaviour of energy variation can be obtained. We consider the maximum energy storage and the faster charging speed (i.e., greater instantaneous power) as indicators to optimize the charging process and find the smallest optimal charging time ${\lambda t_e}$ when the maximum energy obtained by the battery is fixed.}
	
	{Fig. \ref{f2}(a) shows the effect of coupling strength on stored energy with ${\zeta _1} = {\zeta _2} =  \frac{1}{{\sqrt 2 }}$. First, the stored energy increases to its maximum over time. Second, in the long-time limit, the change in the coupling between the qubit and the reservoir cannot increase the peak energy storage; however, it can significantly optimize the shortest time ${\lambda t_s}$ to reach the peak energy storage. Fig. \ref{f2}(b) represents the variation in the energy stored over a larger range of coupling. The energy did not change monotonically with time. Different from Fig. \ref{f2}(a), the enhanced coupling strength enables the QB to store more energy with smaller ${\lambda t_s}$. Fig. \ref{f2}(c) reveals the contribution of coupling strength to instantaneous power during charging. A stronger interaction between the two-qubit system and the reservoir was found to speed up the charging. Compared with Fig. \ref{f2}(c) and Fig. \ref{f2}(d) also reveals that the improvement of the coupling strength is beneficial to the charging power, which is increased by about 4 times. Moreover, the change trend of power does not change like that of energy, that is, it reaches the maximum in a relatively short period of time and then decreases slowly. In short, in Markovian systems, a higher coupling strength reduces the optimal charging time and attains more stored energy; therefore, the charging performance of the battery is improved.}

	\subsection{{Non-Markovian dynamics}}

	{Fig. \ref{f3}(a) clearly shows the help of a memory effect for maintaining high energy storage. The stored energy oscillates efficiently, and the first wave packet represents the maximum stored energy, which corresponds to the optimal charging time. The peaks of the other wave packets decreased with respect to ${\lambda t}$. It can be seen that changing the coupling strength does not significantly improve the maximum stored energy, and a decrease in ${\lambda t_s}$ can be realized by an increase in the coupling. Fig. \ref{f3}(b) illustrates that coupling enhancement can significantly increase the charging speed, which is the same as Fig. \ref{f2}(c) and (d).}
	
	{In the Markovian regime, one can obtain that the maximum energy of the weak coupling regime is about $0.34{\omega _0}$ and the optimal charging time is about 4. However, the non-Markovian system has advantages both in terms of maximum energy storage (the battery is almost full) and optimal charging time $\lambda {t_e} \approx 0.006$, owing to the unique memory effect of the non-Markovian regime, which provides an important way to overcome energy leakage into the environment.}

	{\section{the tightness of an entropic bound and stored energy}}
	
	{In the open quantum battery system, the decoherence effect of the environment leads to an energy leakage. Although greater non-Markovian properties allow the battery to charge satisfactorily for short periods of time, there still exists an additional improvement in energy storage for a longer charging time, which is regarded as another consideration for an excellent charging protocol for quantum batteries. Therefore, in this section, we pursue higher stored energy in the steady state of a quantum battery under the influence of initial entanglement and relative interaction strength ( characterized by comparing the charger-reservoir $\zeta_1$ and battery-reservoir coupling strength $\zeta_2$. Note that $\zeta_1$ and $\zeta_2$ were normalized). Furthermore, we investigate the role of the entanglement and tightness of the entropic bound in the process of boosting stored energy.}
	
	For a system with an X-type density matrix, the concurrence {of the system} \cite{wootters} is $C_{\rho (t)} = 2\max \{ 0,|{\rho _{14}}|  - \sqrt {{\rho _{22}}{\rho _{33}}} ,|{\rho _{23}}|  - \sqrt {{\rho _{11}}{\rho _{44}}} \} $, {where ${\rho _{ij}}$ corresponds to the entries in row $i$ and column $j$ of matrix $\rho (t)$.} Consequently, the concurrence of $\rho _{AB}{\left( t \right)}$ is calculated as ${C_{{\rho _{AB}}(t)}} =2 |{\eta _1}(t){\eta _2}^{*}(t) |$.
	
	To probe the dynamics of entropic uncertainty, we resort to a pair of Pauli operators ${\sigma _x}$ and ${\sigma _z}$ as the incompatibility, which leads to post-measurement states
	\begin{align}
		\rho \left( {{\sigma _x}|B} \right) = \frac{1}{2}\left[ {\rho {{\left( t \right)}_{AB}} + \left( {{\sigma _x} \otimes I} \right)\rho {{\left( t \right)}_{AB}}\left( {{\sigma _x} \otimes I} \right)} \right],
		\label{Eq.15}
	\end{align}
	and
	\begin{align}
		\rho \left( {{\sigma _z}|B} \right) = \frac{1}{2}\left[ {\rho {{\left( t \right)}_{AB}} + \left( {{\sigma _z} \otimes I} \right)\rho {{\left( t \right)}_{AB}}\left( {{\sigma _z} \otimes I} \right)} \right].
		\label{Eq.16}
	\end{align}
	where   $I$ denotes an identity matrix. With respect to $\rho \left( {{\sigma _x}|B} \right)$, {eigenvalues can be calculated} as
	\begin{align}
		\begin{array}{*{20}{c}}
			{\gamma _1^x = \gamma _2^x = \left( {1 - D} \right)/4,} \\
			{\gamma _3^x = \gamma _4^x = \left( {1 + D} \right)/4,}
		\end{array}
		\label{Eq.17}
	\end{align}
	where $D={\sqrt {1 - 4|{\eta _2}\left( t \right){|^2} + 4|{\eta _1}\left( t \right){|^2}|{\eta _2}\left( t \right){|^2} + 4|{\eta _2}\left( t \right){|^4}} }$,
	for $\rho \left( {{\sigma _z}|B} \right)$, we have eigenvalues
	\begin{align}
		\begin{array}{c}
			\gamma _1^z = 0,\gamma _2^z = |{\eta _1}\left( t \right){|^2},\gamma _3^z = |{\eta _2}\left( t \right){|^2},\\
			\gamma _4^z = 1 - |{\eta _1}\left( t \right){|^2} - |{\eta _2}\left( t \right){|^2}.
		\end{array}
		\label{Eq.18}
	\end{align}
	Finally, we obtain the magnitude of the entropic uncertainty as:
	$U_l^{xz}\left( t \right) = 2\sum\limits_{i = 1}^2 {\varepsilon _i^B} {\log _2}\varepsilon _i^B - \sum\limits_{j = 1}^4 {\gamma _j^x} {\log _2}\gamma _j^x - \sum\limits_{j = 1}^4 {\gamma _j^z} {\log _2}\gamma _j^z$ {(left-hand side of Eq. (\ref{Eq.1}))}. {When the system is stable, we have
		\begin{align}
			U_l^{xz}\left( \infty  \right) &= {\left| {{\eta _2}\left( \infty  \right)} \right|^2}{\log _2}{\left| {{\eta _2}\left( \infty  \right)} \right|^2} - M{\log _2}M\nonumber\\
			&- 2\left[ {{{\left| {{\eta _2}\left( \infty  \right)} \right|}^2} - 1} \right]{\log _2}\left[ {1 - {{\left| {{\eta _2}\left( \infty  \right)} \right|}^2}} \right]\nonumber\\
			&- {\left| {{\eta _1}\left( \infty  \right)} \right|^2}{\log _2}{\left| {{\eta _1}\left( \infty  \right)} \right|^2}\nonumber\\
			&+ \left( {F - 1} \right){\log _2}\left[ {\left( {1 - F} \right)/4} \right]/2\nonumber\\
			&- \left( {F + 1} \right){\log _2}\left[ {\left( {F + 1} \right)/4} \right]/2,
			\label{Eq.19}
	\end{align}}
and lower bound $U_r^2\left( t \right) = 1 + \sum\limits_{i = 1}^2 {\varepsilon _i^B} {\log _2}\varepsilon _i^B - \sum\limits_{j = 1}^4 {\varepsilon _j^{AB}} {\log _2}\varepsilon _j^{AB}$ {( right-hand side of Eq. ~ (\ref{Eq.1}))}. {The corresponding lower bound in the steady state is in the form of
		\begin{align}
			U_r^2\left( \infty  \right) &= 1 + {\left| {{\eta _2}\left( \infty  \right)} \right|^2}{\log _2}{\left| {{\eta _2}\left( \infty  \right)} \right|^2}\nonumber\\
			&- \left[ {{{\left| {{\eta _2}\left( \infty  \right)} \right|}^2} - 1} \right]{\log _2}\left[ {1 - {{\left| {{\eta _2}\left( \infty  \right)} \right|}^2}} \right]\nonumber\\
			&- M{\log _2}M - \left( {1-M} \right){\log _2}\left( {1-M} \right),
			\label{Eq.20}
		\end{align}
		where ${\varepsilon _i^B}$ and ${\varepsilon _j^{AB}}$ are the eigenvalues of $\rho _B{\left( t \right)}$ and $\rho_{AB} {\left( t \right)}$, $F = \sqrt {1 + 4{{\left| {{\eta _1}\left( \infty  \right)} \right|}^2}{{\left| {{\eta _2}\left( \infty  \right)} \right|}^2} - 4{{\left| {{\eta _2}\left( \infty  \right)} \right|}^2} + 4{{\left| {{\eta _2}\left( \infty  \right)} \right|}^4}}  $, and $M = 1 - {\left| {{\eta _1}\left( \infty  \right)} \right|^2} - {\left| {{\eta _2}\left( \infty  \right)} \right|^2}$. The superscripts of entropic uncertainty and lower bounds indicate the Pauli operator chosen for the measurement and the number of measurement operators, respectively.}
	
	\begin{figure*}[htbp]
		\centering
		\subfigure{\includegraphics[height=4.5cm]{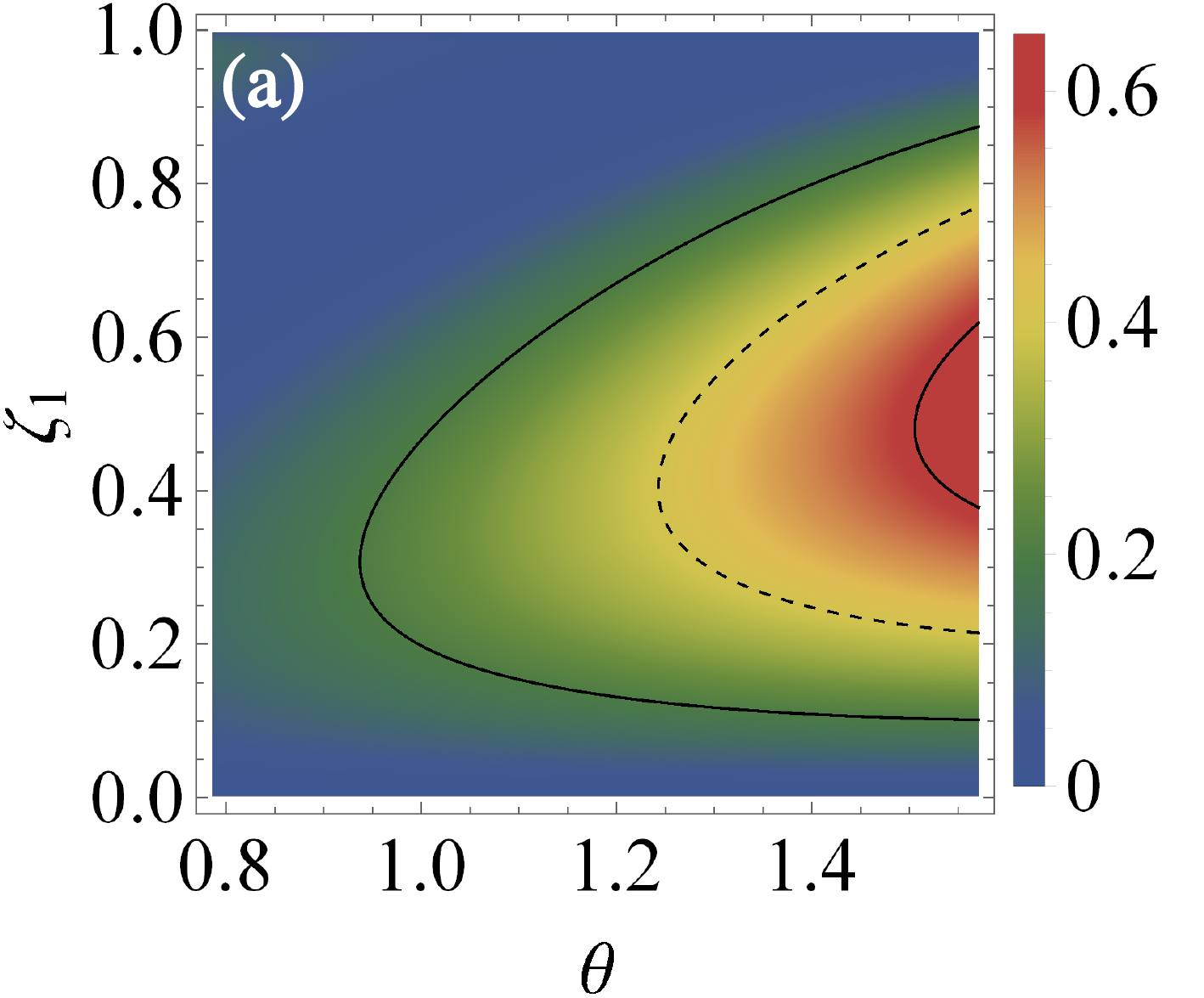}}
		\hspace{0.5cm}
		\subfigure{\includegraphics[height=4.5cm]{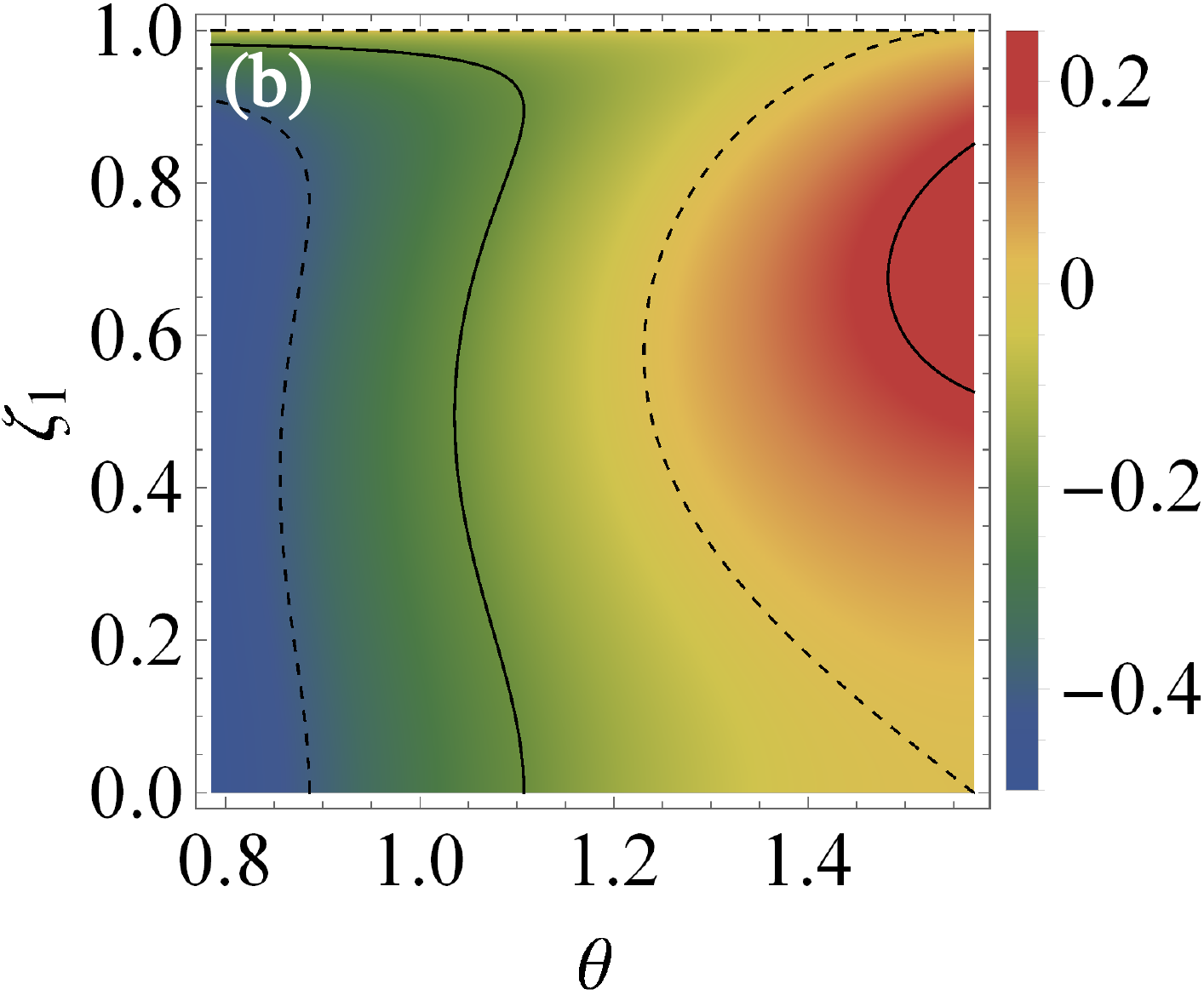}}
		\hspace{0.5cm}
		\subfigure{\includegraphics[height=4.5cm]{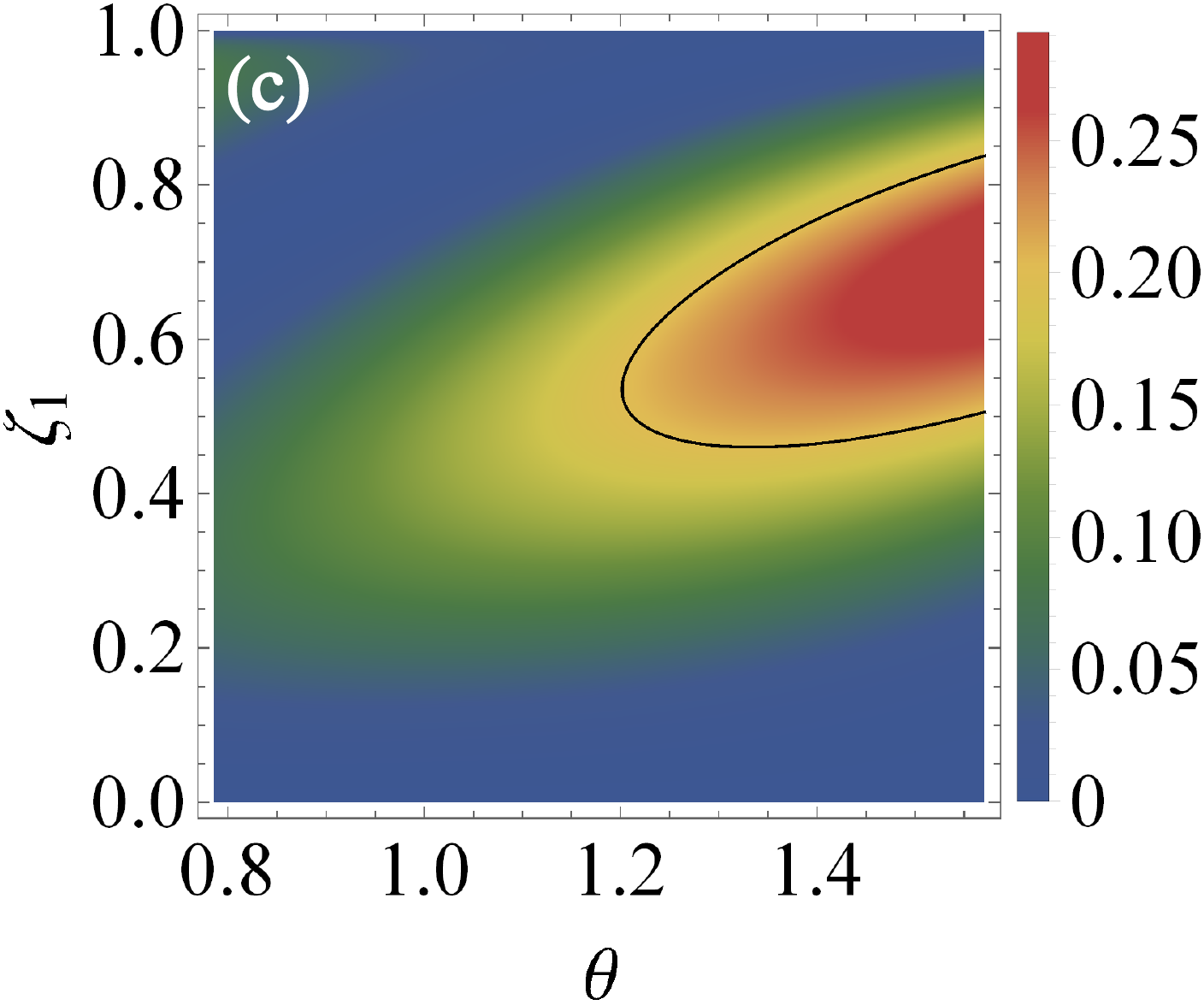}}
		\\
		\subfigure{\includegraphics[height=4.5cm]{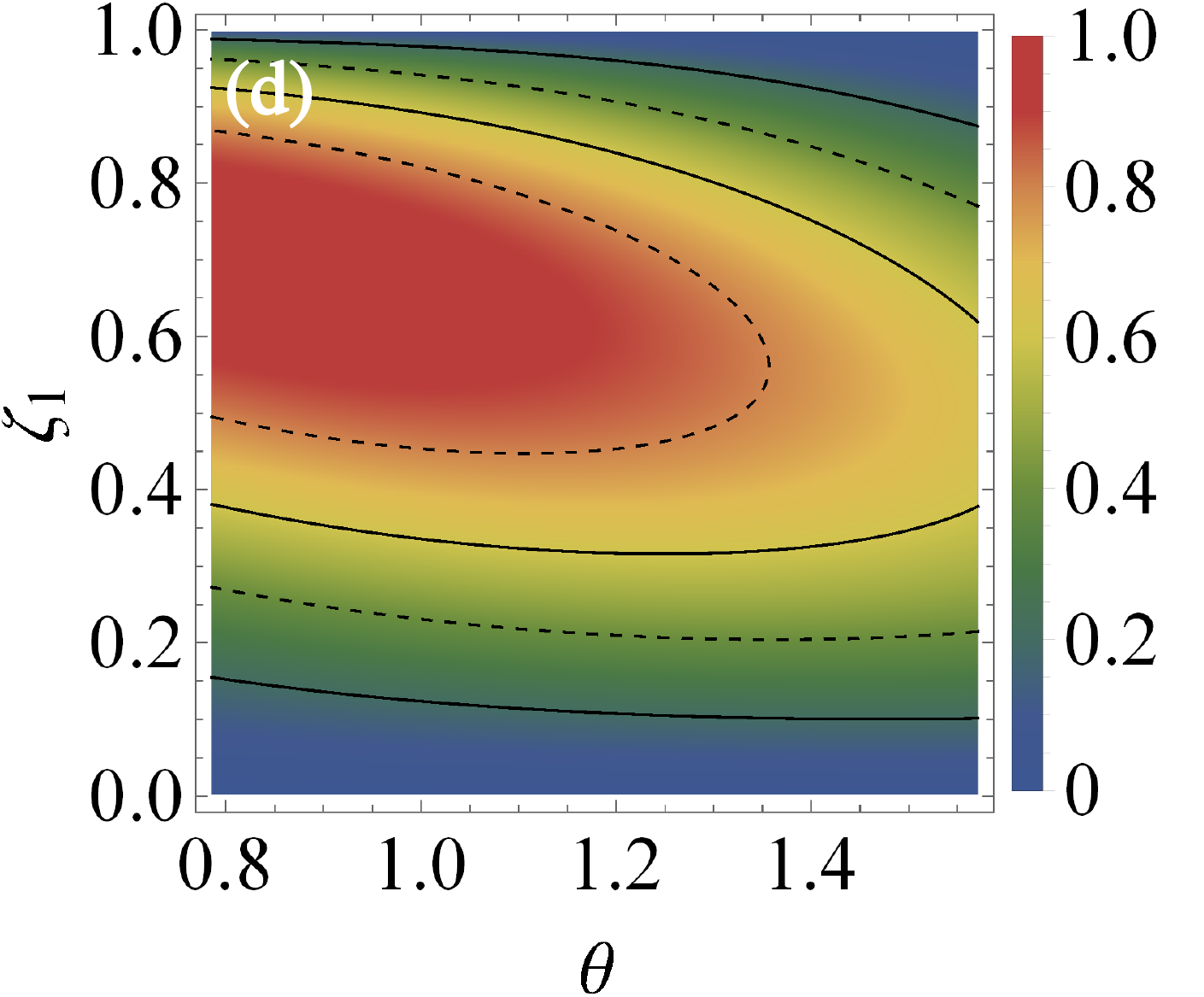}}
		\hspace{0.5cm}
		\subfigure{\includegraphics[height=4.5cm]{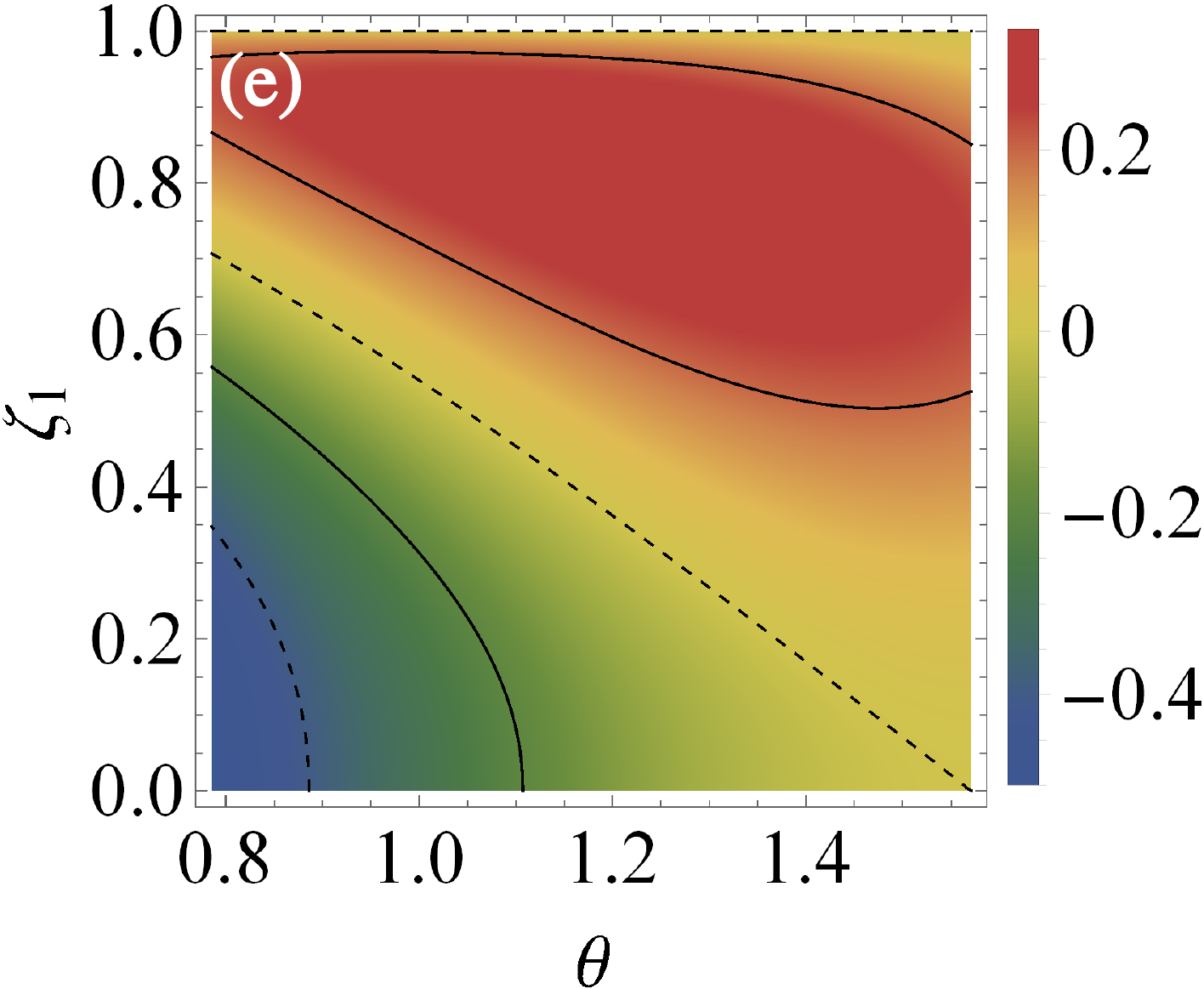}}
		\hspace{0.5cm}
		\subfigure{\includegraphics[height=4.5cm]{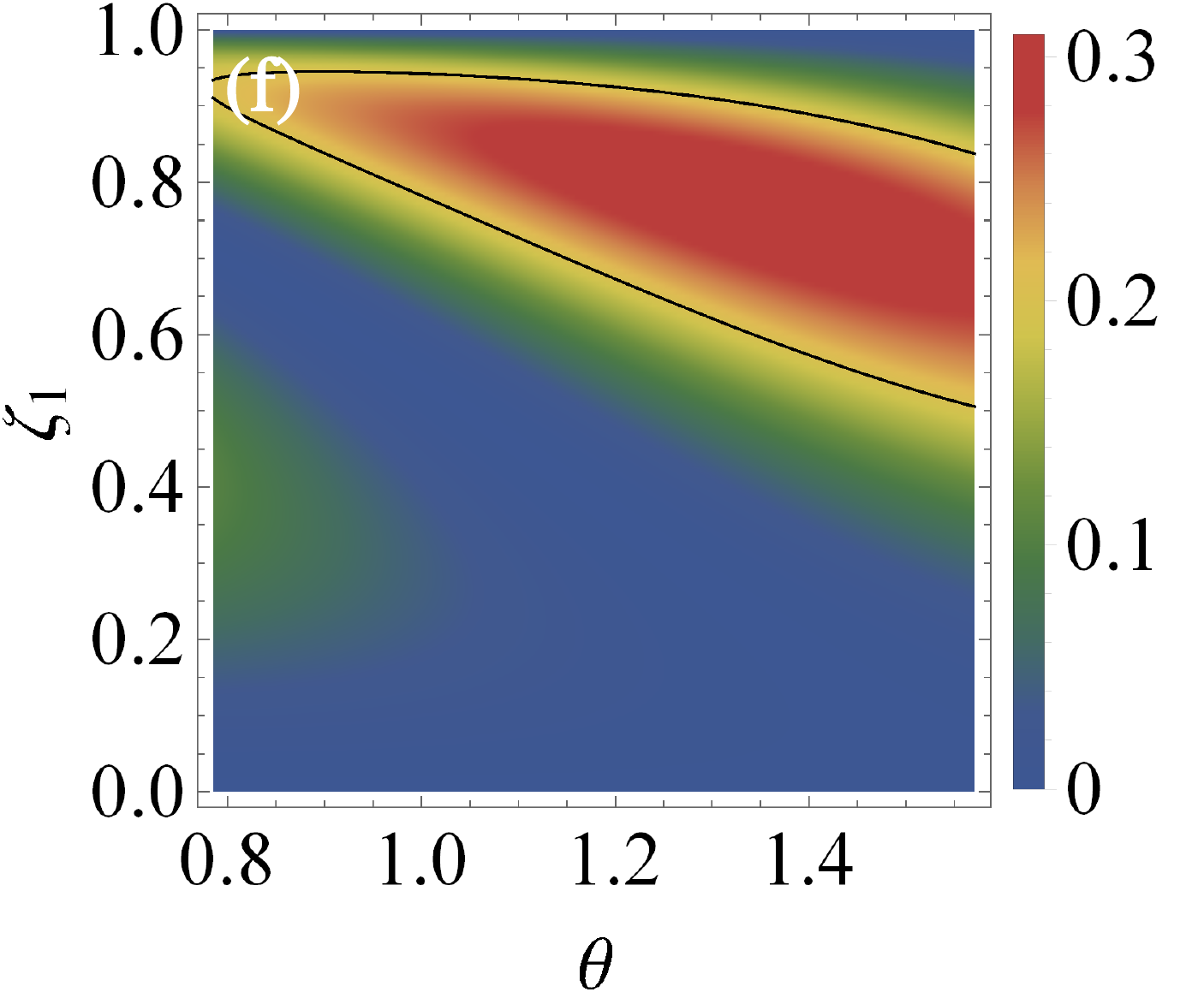}}
		\caption{{When the battery is in the steady-state,  the concurrence ${C_{{\rho _{AB}}(\infty )}}$, stored energy, $\Delta {E_B}(\infty )/{\omega _0}$ and the tightness of the lower bound $\Delta U^{xz}(\infty )$ are affected by the relative interaction strength ${\zeta_i}$ and the probability amplitude of the initial state. ${C_{{\rho _{AB}}(\infty )}}$ is plotted in (a) and (d); $\Delta {E_B}(\infty )/{\omega _0}$ is plotted in (b) and (e); $\Delta U^{xz}(\infty )$ is plotted in (c) and (f); $\theta  \in [0.25\pi,0.5\pi ]$ and ${\zeta _1} \in [0,1]$ are given, respectively. In addition, the relative phase is also associated with  the concurrence ${C_{{\rho _{AB}}(\infty )}}$, stored energy, $\Delta {E_B}(\infty )/{\omega _0}$ and the tightness of the lower bound $\Delta U^{xz}(\infty )$, here, we set  $\phi=0$ in Graphs (a), (b), and (c); and $\phi=\pi$ in Graphs (d), (e) and (f).}}
		\label{f4}
	\end{figure*}
	{After a sufficient charging period, i.e., $t \to \infty $, the internal energy of the battery tends to be stable, and at this time we get $\mu (\infty ) = 0$,  ${\eta _1}(\infty ) = \zeta _2^2{\eta _{01}} - {\zeta _1}{\zeta _2}{\eta _{02}}$, and ${\eta _2}(\infty ) = \zeta _1^2{\eta _{02}} - {\zeta _1}{\zeta _2}{\eta _{01}}$. Thus, the
		The concurrence can be expressed as follows:
		\begin{align}
			{C_{{\rho _{AB}}(\infty )}} = 2|{\eta _1}(\infty )  \eta _2^*(\infty )|,
			\label{Eq.21}
		\end{align}
		and the stored energy of the battery in the steady state is obtained by:
		\begin{align}
			\Delta {E_B}\left( \infty  \right) = {\omega _0}\left[ |{\eta _2}\left( \infty  \right){|^2} - \cos ^2  \theta \right].
			\label{Eq.22}
		\end{align}
		Similarly, entropic uncertainty ${U_l}^{xz}(\infty )$ and its lower bound ${U_{r}^{2}}(\infty )$ in the steady state can be calculated. The tightness of the lower bound is given by $\Delta U^{xz}(\infty ) = {U_l}^{xz}(\infty ) - {U_{r}^{2}}(\infty )$, which reflects the general characteristics of the entropic uncertainty relation. The tighter lower bound is quantified by a smaller $\Delta U^{xz}(\infty )$. }
	
	{In addition, as a parameter in the probability amplitude of the initial state, $\theta$ is reflected in the initial charger-battery entanglement in the charging scheme. The concurrent initial state can be expressed as ${C_{{\rho _{AB}}(0)}} = 2|{\eta _{01}}\eta _{02}^*|$. After the calculation, it was found that the entanglement of the initial state is only related to $\theta $ and is independent of the phase $\phi$, which reaches a maximum value of $1$ when $\theta =0.25\pi$ and monotonically decreases to $0$ when $\theta $ approaches the boundary value $0.5\pi$. Therefore, in the charging process, we consider the influence of the initial entanglement on battery performance according to the change in $\theta $. }
	
	{In Fig. \ref{f4}(a), (b), and (c), entanglement and stored energy have a similar evolution process while they are opposite to the tightness of entropic bound. All three quantities reach an extreme value when the initial entanglement is at a minimum. Moreover, when the initial state is separable ($\theta=0.5\pi$), the stored energy is effective ($\Delta {E_B}(\infty ) \ge 0$) for all values of the relative interaction strength, and the stored energy reaches a maximum ($\Delta {E_B}{(\infty )_{\max }} = 0.25{\omega _0}$) when the coupling strengths of the charger-reservoir  and the battery-reservoir are symmetrical (${\zeta _1} = {\zeta _2} = {1 \mathord{\left/
				{\vphantom {1 {\sqrt 2 }}} \right.
				\kern-\nulldelimiterspace} {\sqrt 2 }}$). The results also show that entanglement is beneficial for energy storage in when $\phi=0$. Although a tighter entropic bound is not required for the growth of stored energy, there is a close relationship between them, that is, the highest stored energy strictly corresponds to the least tight entropic bound, which never occurs with respect to entanglement.}
	
	{In Fig. \ref{f4}(d), (e), and (f), with the reversal of the relative phase to $\phi  = \pi $, it can be seen that entanglement reaches the peak value when the system is initially the maximum entangled state, which is completely opposite to that of the case of $\phi  = 0 $. Simultaneously, the entropic bound became less tight. Furthermore, there is effective stored energy ($\Delta {E_B}(\infty ) \ge 0$), regardless of the initial charger-battery entanglement. Compared to the case of $\phi  = 0 $, the peak value of the stored energy increases to approximately $0.33{\omega _0}$ when $\phi  = \pi $, accompanied by an asymmetric charging process (${\zeta _1} \approx 0.8$) and a certain initial entanglement ($\theta  \approx 1.3$). Comparing Fig. \ref{f4}(d) and (e), it is find that charger-battery entanglement is necessary in the parameter plane $\theta-{\zeta _1}$ to obtain the most stored energy. Interestingly, the flip of the relative phase does not change the close relationship between the tightness of the entropic bound and the stored energy, that is, the most stored energy when $\phi  = \pi $ strictly corresponds to the least tight entropic bound. }
	\begin{figure*}
		\begin{minipage}{1\textwidth}
			\centering
			\subfigure{\includegraphics[width=7.5cm]{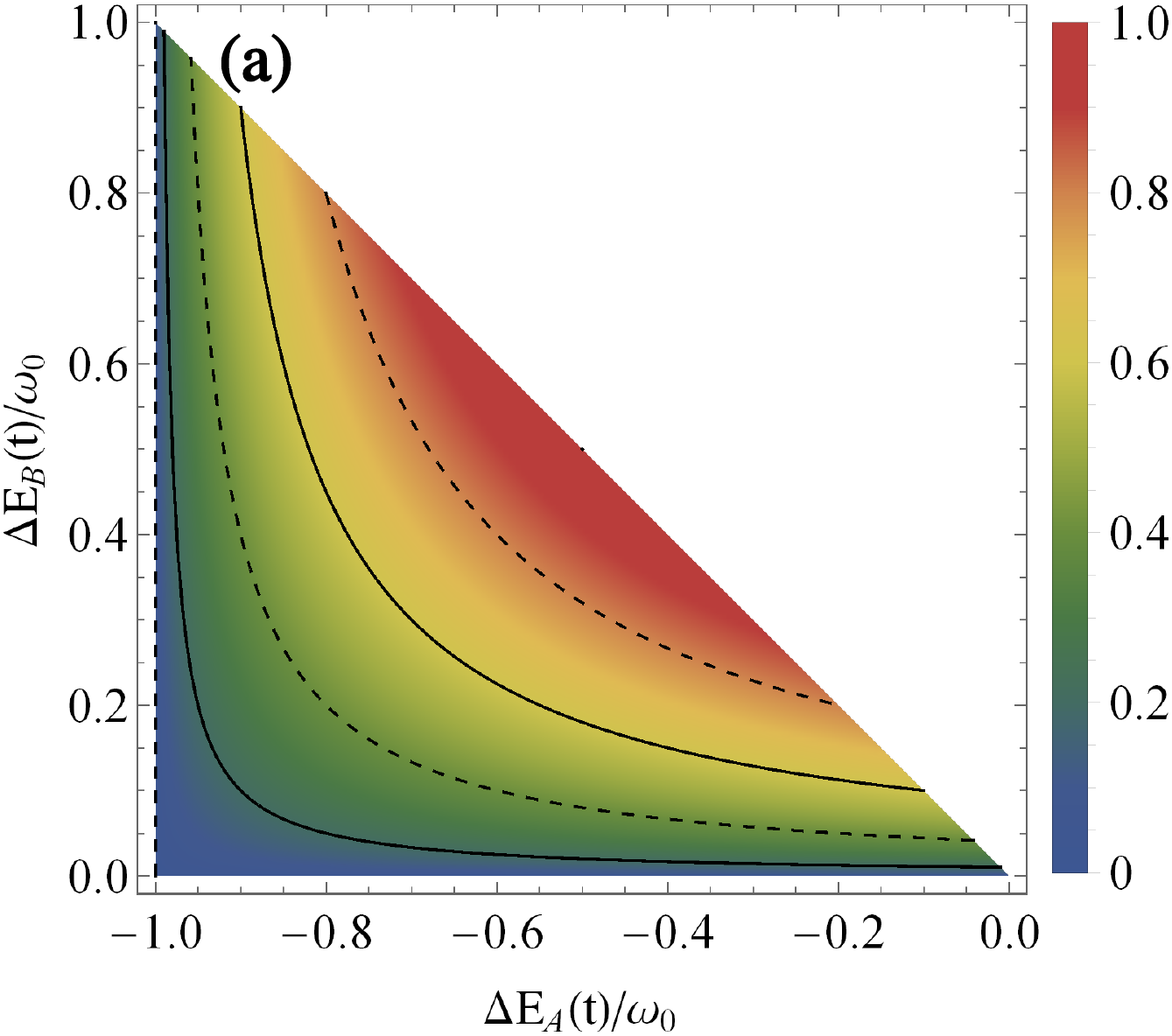}} \ \ \ \ \
			\subfigure{\includegraphics[width=7.5cm]{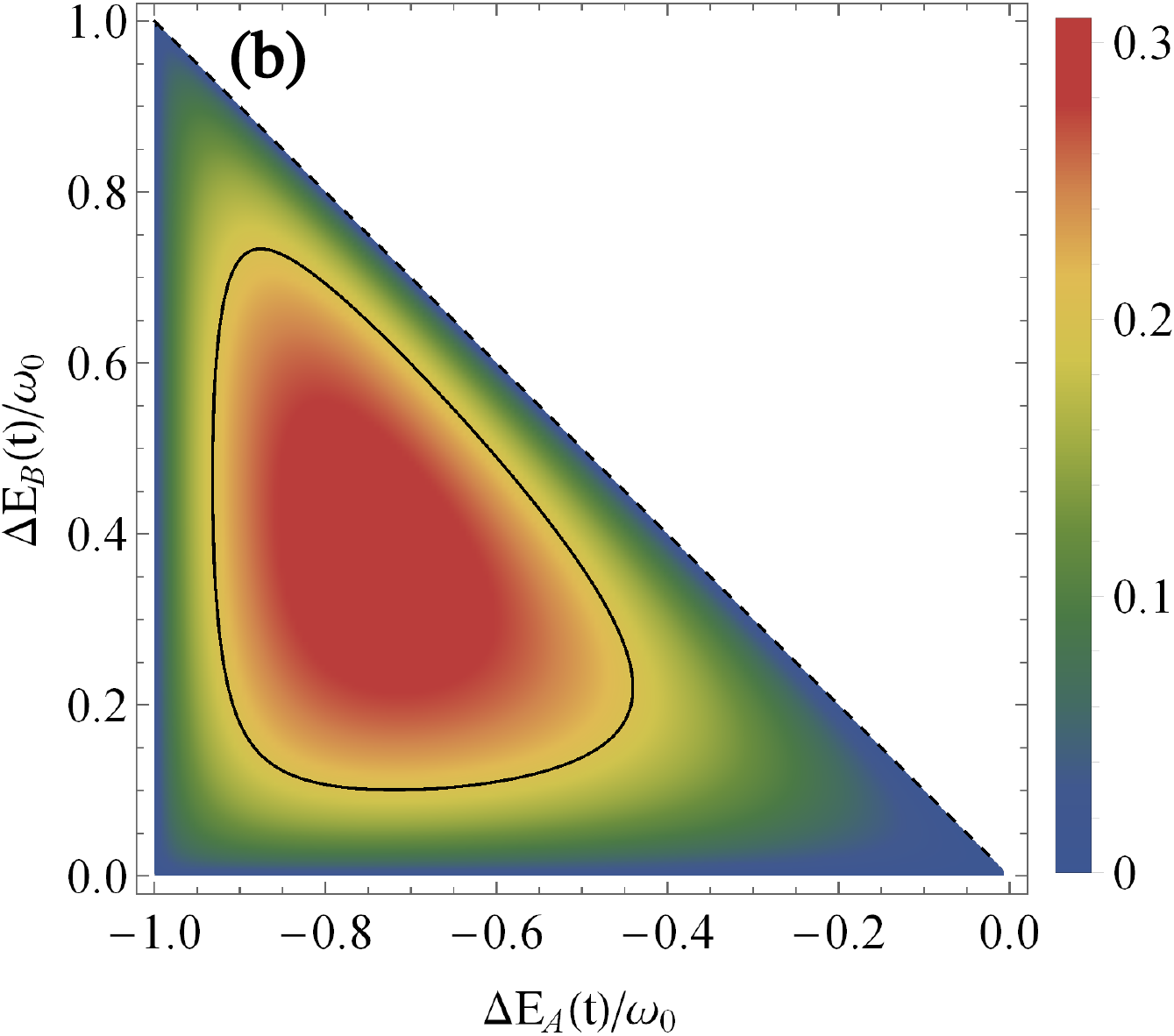}}
		\end{minipage}\hfill
		\caption{{The concurrence ${C_{{\rho _{AB}}(t)}}$ and the tightness $\Delta U^{xz}(t)$ as functions of energy are plotted in (a) and (b), respectively. The x-axis is the energy change of the charger $\Delta {E_A}(t)/{\omega _0}$, and the y-axis is the stored energy of the battery $\Delta {E_B}(t)/{\omega _0}$. In addition, a separable initial state $\theta  = 0.5\pi $ is set.}}
		\label{f5}
	\end{figure*}
	{We find that the stored energy of the steady state is always negatively correlated with the tightness of the entropic bound when the relative phase is $0$ or $\pi $. Therefore, we will further study the relationship between the stored energy and tightness of the entropic bound from the transfer of energy between charger and battery. Here, we study the evolution of the entanglement and tightness of the entropic bound for any case of energy transfer. The energy variation of the charger and battery is associated with the tightness of the lower bound and the entanglement (according to Eqs. (\ref{Eq.19}), (\ref{Eq.20}), (\ref{Eq.21}), and (\ref{Eq.22})). From the relation ${\left| {{\eta _1}\left( t \right)} \right|^2} = \Delta {E_A}(t)/{\omega _0} + \sin^{2} {\theta }$ and ${\left| {{\eta _2}\left( t \right)} \right|^2}= \Delta {E_B}(t)/{\omega _0} + \cos^{2} {\theta }$, where, $\left| {\Delta {E_A}\left( t \right)} \right| $ is the energy lost by the charger, we can see that the energy partly determines the evolutionary behavior of entanglement or tightness. The concurrence can then be expressed as ${C_{{\rho _{AB}}\left( t \right)}} = 2{\left[ {\Delta {E_A}\left( t \right)/{\omega _0} + \sin^{2} {{ \theta  }}} \right]^{\frac{1}{2}}}{\left[ {\Delta {E_B}\left( t \right)/{\omega _0} + \cos^{2} {{ \theta  }}} \right]^{\frac{1}{2}}}$. Similarly, the tightness of the entropic bound can be represented by $\Delta {U^{xz}}\left( t \right)$, which is obtained by substituting energy and parameters $\theta$.}
	
	{Next, we visually depict the relationship between entanglement or tightness and energy storage in  Fig. \ref{f5}.
		The energy stored by the battery during the charging process has a boundary value, which is the maximum value of the stored energy regardless of the energy lost by the charger, as shown in Fig. \ref{f5} as the hypotenuse of a triangle region. The boundary value or constraint can be expressed as $\Delta {E_B}(t)/|\Delta {E_A}(t)| = 1$. This implies that the energy lost by the charger is completely transferred to the battery. However, we find that at $\Delta {E_B}(t) \ne |\Delta {E_A}(t)|$, that is, $\Delta {E_B}\left( t \right)/|\Delta {E_A}\left( t \right)| < 1$ the energy is not completely transferred to the battery, which means that some of the transferred energy flows into the environment. Different energy transfers actually correspond to arbitrary charging processes caused by different parameter settings in the charging model. Therefore, it is meaningful to discuss the effect of energy transfer rate $\Delta {E_B}\left( t \right)/|\Delta {E_A}\left( t \right)|$ on the tightness of entropic bound.}
	
	{Meanwhile, Fig. \ref{f5}(a) shows that the relationship between entanglement and energy variation during the arbitrary charging process. It can be seen that an increase in the energy transfer rate (i.e., the charger loses less energy or the battery stores more energy) contributes to an increase in entanglement. Entanglement disappears when the battery or charger is empty.} {Interestingly, the energy can be completely transferred for different entanglements, corresponding to the states of the hypotenuse in Fig. \ref{f5}(a). Therefore, we can conclude that entanglement is not  a clear indicator of the maximal energy stored.}
	{Besides, we obtain that  the maximum energy transfer rate corresponds to the tightest entropic bound except when the charger or battery is completely empty, as shown in  Fig. \ref{f5}(b). This indicates that the tightness of the entropic bound can also play a role in indicating the degree of energy transfer in  the charging processes, implying that the energy is completely transferred for  the tightest  entropic bound $\Delta {U^{xz}}\left( t \right) = 0$, which virtually agrees with the previous conclusion.}

	{To explore the role of tightness in improving charging in a more general context. We consider three measurements with $(\sigma_x,\sigma_y,\sigma_z)$.}
	{As a result, the entropic uncertainty is
		\begin{align}
			U_l^{xyz}\left( t \right) &=  - {\left| {{\eta _1}\left( t \right)} \right|^2}{\log _2}{\left| {{\eta _1}\left( t \right)} \right|^2} + 2{\left| {{\eta _2}\left( t \right)} \right|^2}{\log _2}{\left| {{\eta _2}\left( t \right)} \right|^2} \nonumber \\ \nonumber
			&- 3\left[ {{{\left| {{\eta _2}\left( t \right)} \right|}^2} - 1} \right]{\log _2}\left[ {1 - {{\left| {{\eta _2}\left( t \right)} \right|}^2}} \right] - m{\log _2}m \\
			&+ \left( {D - 1} \right){\log _2}\left[ {\left( {1 - D} \right)/4} \right] \nonumber\\
			&- \left( {D + 1} \right){\log _2}\left[ {\left( {1 + D} \right)/4} \right],
			\label{Eq.23}
		\end{align}
		where $m = 1 - {\left| {{\eta _1}\left( t \right)} \right|^2} - {\left| {{\eta _2}\left( t \right)} \right|^2}$ and the corresponding lower bound is
		\begin{align}
			{U_{r}^{3}}\left( t \right) = 2{U_{r}^{2}}\left( t \right) - 1
			\label{Eq.24},
		\end{align}
		the tightness of the three measurements was $\Delta U^{xyz}(t)= U_{l}^{xyz}(t)- U_{r}^{3}(t)$.}
	\begin{figure}
		\begin{minipage}{0.5\textwidth}
			\centering
			\subfigure{\includegraphics[width=7cm]{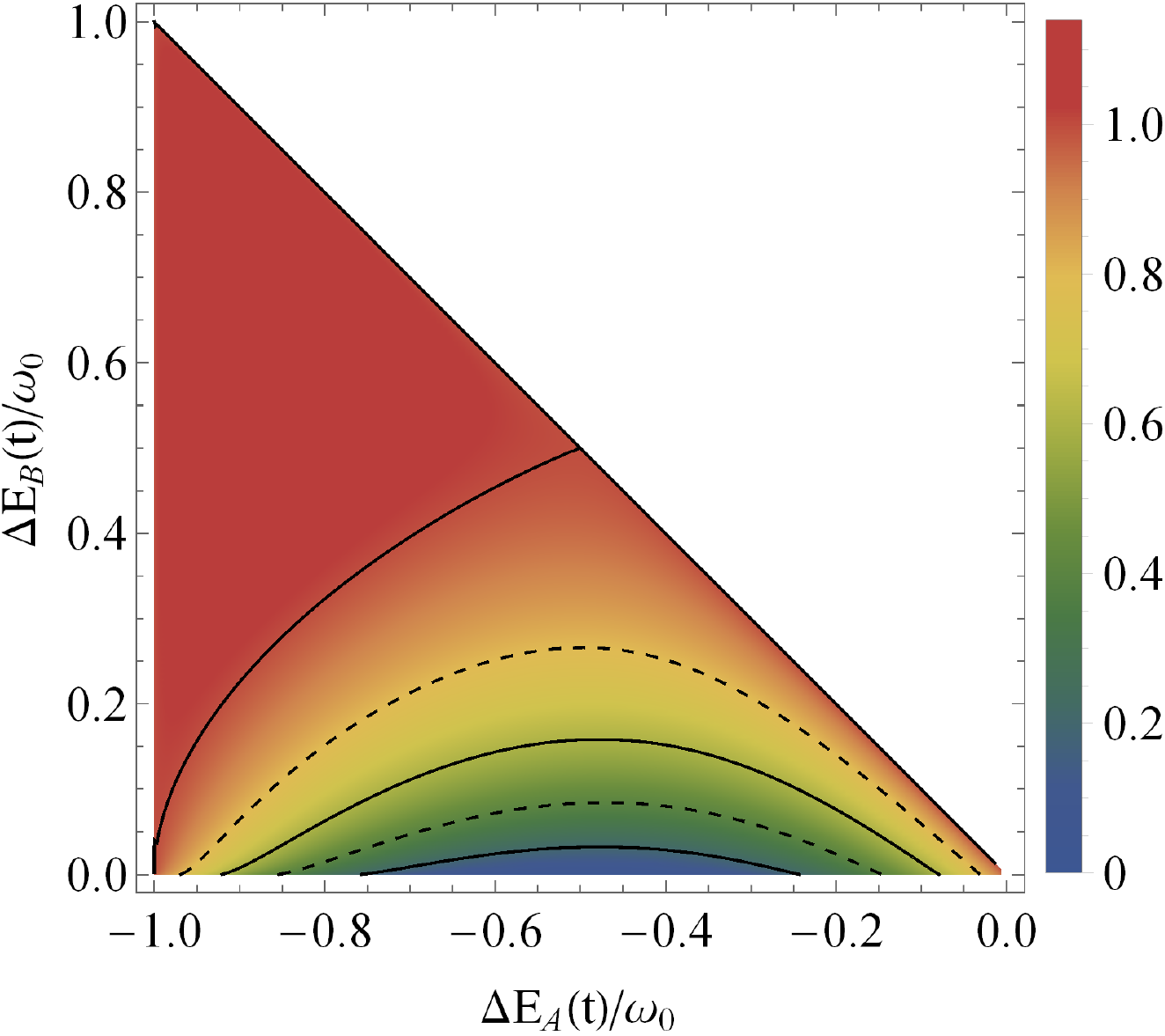}}
		\end{minipage}\hfill
		\caption{{Tightness $\Delta U^{xyz}(t)$ as functions of energy are plotted. The other settings are the same as those in Fig. \ref{f5}.}}
		\label{f6}
	\end{figure}
	{Fig. \ref{f6} shows the connection between the exchanged energy  and the  tightness of the entropic uncertainty.}
	{Obviously, the complete energy transfer takes place when the tightness is fixed to $\Delta {U^{xyz}}\left( t \right) = 1$ corresponding to the hypotenuse, which also supports
		tightness can be considered an effective indicator of the maximum energy transfer during the charging of QBs.}

	\section{Discussions and conclusions}
	{In summary, we have investigated the dynamics of the entropic uncertainty of charging quantum batteries in common dissipative bosonic environments. We studied two cases of the charging process, and the results show that the coupling enhancement during the Markovian charging process cannot only improve the charging power, but also increase the stored energy of the battery. Stronger coupling in non-Markovian systems can also significantly improve the charging power; in contrast, the battery can be efficiently full during the non-Markovian charging process. When the battery is in a steady state, maximum energy storage is achieved when the charger is more coupled to the reservoir than the battery, in the case of a certain initial charger-battery entanglement. It is worth mentioning that a tighter entropic bound has a negative effect on energy storage in the presence of entanglement. We then studied the dynamic behavior of the entanglement and tightness in an arbitrary charging process, and the results revealed that an increase in entanglement can increase the energy transfer rate and degrade the energy storage. Remarkably, it is argued that the tightness of the entropic bound can be regarded as an important indicator of the optimal energy transfer during charging, and the tightest entropic bound corresponds to the complete energy transfer. We believe that our findings will be helpful for the pursuit of high-performance energy transfer in $N$-cell QB charging in the future.}

	\begin{acknowledgements}
		This study was supported by the National Natural Science Foundation of China (Grant Nos. 12075001, 61601002, and 12175001).
		Anhui Provincial Key Research and Development Plan (Grant No. 2022b13020004), Anhui Provincial Natural Science Foundation (Grant No. 1508085QF139), and Fund of the CAS  Key  Laboratory  of
		Quantum  Information (Grant No. KQI201701).
	\end{acknowledgements}

\end{document}